\documentclass[proof]{pasj00}

\begin{document}
\SetRunningHead{S. Nakahira et al.}{MAXI/GSC observations of XTE J1752--223}
%\Received{}%{yyyy/mm/dd}
%\Accepted{}%{yyyy/mm/dd}
%\Published{}%{yyyy/mm/dd}

\title{MAXI GSC observations of a spectral state transition \\ in the black hole candidate XTE J1752--223}

%%% begin:list of authors
% Do NOT capitalize all letters in "textsc".
\author{Satoshi Nakahira\altaffilmark{1}, Kazutaka Yamaoka\altaffilmark{1}, Mutsumi Sugizaki\altaffilmark{2}, 
Yoshihiro Ueda\altaffilmark{3}, Hitoshi Negoro\altaffilmark{4}, Ken Ebisawa\altaffilmark{5}, Nobuyuki Kawai\altaffilmark{6,2}, 
Masaru Matsuoka\altaffilmark{2,7}, Hiroshi Tsunemi\altaffilmark{8}, Arata Daikyuji\altaffilmark{9}, Satoshi Eguchi\altaffilmark{3}, 
Kazuo Hiroi\altaffilmark{3}, Masaki Ishikawa\altaffilmark{10}, Ryoji Ishiwata\altaffilmark{4}, Naoki Isobe\altaffilmark{3}, 
Kazuyoshi Kawasaki\altaffilmark{7}, Masashi Kimura\altaffilmark{8}, Mitsuhiro Kohama\altaffilmark{2,7}, 
Tatehiro Mihara\altaffilmark{2}, Sho Miyoshi\altaffilmark{4}, Mikio Morii\altaffilmark{6}, Yujin E. Nakagawa\altaffilmark{11}, 
Motoki Nakajima\altaffilmark{12}, Hiroshi Ozawa\altaffilmark{4}, Tetsuya Sootome\altaffilmark{2}, Kousuke Sugimori\altaffilmark{6}, 
Motoko Suzuki\altaffilmark{2}, Hiroshi Tomida\altaffilmark{7}, Shiro Ueno\altaffilmark{7}, Takayuki Yamamoto\altaffilmark{2}, 
Atsumasa Yoshida\altaffilmark{1,2}, and the MAXI team.}

\altaffiltext{1}{Department of Physics and Mathematics, Aoyama Gakuin University, 5-10-1 Fuchinobe, Chuo-ku, Sagamihara, Kanagawa 252-5258, Japan}
\email{nakahira@phys.aoyama.ac.jp}
\altaffiltext{2}{Coordinated Space Observation and Experiment Research Group, Institute of Physical and Chemical Research (RIKEN), 2-1 Hirosawa, Wako, Saitama 351-0198, Japan}
\altaffiltext{3}{Department of Astronomy, Kyoto University, Oiwake-cho, Sakyo-ku, Kyoto 606-8502, Japan}
\altaffiltext{4}{Department of Physics, Nihon University, 1-8-14 Kanda-Surugadai, Chiyoda-ku, Tokyo 101-8308, Japan}
\altaffiltext{5}{Department of Space Science Information Analysis, Institute of Space and Astronautical Science, Japan Aerospace Exploration Agency , 3-1-1 Yoshino-dai, Chuo-ku, Sagamihara, Kanagawa 252-5210, Japan}
\altaffiltext{6}{Department of Physics, Tokyo Institute of Technology, 2-12-1 Ookayama, Meguro-ku, Tokyo 152-8551, Japan}
\altaffiltext{7}{ISS Science Project Office, Institute of Space and Astronautical Science, Japan Aerospace Exploration Agency, 2-1-1 Sengen, Tsukuba, Ibaraki 305-8505, Japan}
\altaffiltext{8}{Department of Earth and Space Science, Osaka University, 1-1 Machikaneyama, Toyonaka, Osaka 560-0043, Japan}
\altaffiltext{9}{Department of Applied Physics, University of Miyazaki, 1-1 Gakuen Kibanadai-nishi, Miyazaki, Miyazaki 889-2192, Japan}
\altaffiltext{10}{School of Physical Science, Space and Astronautical Science, The graduate University for Advanced Studies (Sokendai), Yoshinodai 3-1-1, Chuo-ku, Sagamihara, Kanagawa 252-5210, Japan}
\altaffiltext{11}{High Energy Astrophysics Laboratory, Institute of Physical and Chemical Research (RIKEN), 2-1 Hirosawa, Wako, Saitama 351-0198, Japan}
\altaffiltext{12}{School of Dentistry at Matsudo, Nihon University, 2-870-1 Sakaecho-nishi, Matsudo, Chiba 101-8308, Japan}

%\author{C-Firstname {\sc C-Familyname}}
%\affil{C-Address of Institute}\email{ccccc@xxx.xxx.xx.xx}
%%% end:list of authors

%%% Please use the following style in case that sorting by 
%%% affilation is impossible. 
%
% \author{%
%   D-Firstname \textsc{D-Familyname}\altaffilmark{1}
%   E-Firstname \textsc{E-Familyname}\altaffilmark{1,2}
%   and
%   F-Firstname \textsc{F-Familyname}\altaffilmark{2}}
% \altaffiltext{1}{Address of Institute}
% \email{ddddd@xxx.xxx.xx.xx}
% \email{eeeee@xxx.xxx.xx.xx}
% \altaffiltext{2}{Address of Institute}

%% `\KeyWords{}' always has to be placed before `\maketitle'.
 %Do NOT move this preamble from here!

\maketitle
\begin{abstract}

 We present the first results on the black hole candidate XTE J1752--223 
 from the Gas Slit Camera (GSC) on-board the Monitor of All-sky X-ray Image (MAXI) 
 on the International Space Station. 
Including the onset of the outburst reported by the Proportional Counter Array on-board
  the Rossi X-ray Timing Explorer on 2009 October 23, the MAXI/GSC has 
  been monitoring this source approximately 10 times per day with a high sensitivity in the 2--20 keV band. 
 XTE J1752--223 was initially in the low/hard state during the first 3 months.
 An anti-correlated behavior between the 2--4 keV and 4--20 keV bands were observed around January 20, 2010,  
  indicating that the source exhibited the spectral transition to the high/soft state. 
 A transient radio jet may have been ejected when the source was in the intermediate state where 
 the spectrum was roughly explained by a power-law with a photon
 index of 2.5--3.0. The unusually long period in the initial low/hard state implies a slow variation 
 in the mass accretion rate, and the dramatic soft X-ray increase may be explained by a sudden 
 appearance of the accretion disk component with a relatively low innermost temperature (0.4--0.7 keV). 
 Such a low temperature might suggest that the maximum accretion rate was 
 just above the critical gas evaporation rate required for the state transition.

\end{abstract}
% \linenumbers　
\section{Introduction}

Galactic Black hole candidates (BHCs) are ideal objects in studying how accretion disks 
 and coronae evolve due to changes in the accretion rate, thanks to their
  high photon statistics and frequent state transitions (see, e.g.,  \citet{review} for a  review).
They are characterized by transient behavior, such as a sudden X-ray brightening 
caused by instabilities originating in the outer 
accretion disk. During the outburst, which typically lasts a few months, 
  they may go through  several distinct spectral states,  initial low/hard state (LHS), intermediate or 
 very high state (IMS/VHS), high/soft state (HSS), LHS, then go back to quiescence. 
These  states are believed to reflect changes in the geometry of the accretion disk and corona (\cite{flow1}), as well as  
 jet (\cite{flow2}, \cite{flow3}). BHCs sometimes show powerful radio jets in their outbursts 
 in association with state transitions from the LHS to the HSS. Hence, study of the spectral transition and the disk-jet
 interaction is important in clarifying jet production mechanisms  (see, e.g.,  \citet{review_jet}  for a review).  

The new X-ray transient XTE J1752--223 was first detected at the 30 mCrab flux level 
 in the 2--10 keV range on October 23, 2009 during the galactic bulge monitoring with the Proportional Counter 
 Array (PCA) of the Rossi X-ray Timing Explorer (RXTE) \cite{discovery}.
 On October 24, the source also triggered the Swift Burst Alert Telescope (BAT) at a 
  100 mCrab flux in the 15--50 keV range \citep{discovery}. The position was determined to 
 (RA,Dec)=(17$^{\rm h}$52$^{\rm m}$15$^{\rm s}$.14, --22$^{\circ}$20'33''.8) with  5 arc-sec uncertainty 
  by a  follow-up Swift X-Ray Telescope (XRT) localization \citep{swiftxrt}.
 The energy spectrum is consistent with a power-law with photon index  1.38$\pm$0.01 and 
   the power spectrum shows band-limited noise with a total rms of $\sim$30 percent, suggesting 
 that the source was in a typical LHS (\cite{xte_spec}; \cite{xte_result}). 
 The optical \citep{opt}, near-infrared \citep{opt+nir}, and radio counterparts \citep{radio} 
 were also discovered within the X-ray error circle during the initial phase. 
 Possible double-sided jets were detected on February 11, 2010 (\cite{jet}; \cite{jet2}); 
 however, the BH mass, inclination angle, and distance have not been determined yet.

Since August 3, 2009, the highly sensitive all-sky monitor, Monitor of All-sky X-ray Image, 
MAXI(\cite{maxi}), has been activated on the Exposed Facility of the Japanese Experimental 
Module ``Kibo'' of the International Space Station (ISS). Soon after the discovery of XTE J1752--223, 
 when the MAXI was still in the commissioning phase, the
 source  was found at the 30 mCrab level in an X-ray image of the MAXI Gas Slit Camera (GSC) (\cite{maxiatel1}). 
 MAXI/GSC also observed a rapid spectral softening on 2010 January 22 due to 
 a spectral transition from the LHS to the HSS (\cite{maxiatel2}).
The MAXI initial results are described in \citet{maxi_first} and \citet{maxi_first2}.

 In this paper, we report on  MAXI GSC observations of XTE J1752--223 during 
 its outburst since October 2009. In the observation and result section,  we 
demonstrate  MAXI's spectral performance for such a  bright source. 
Finally, we discuss the origin of the rather unusual behavior 
of XTE J1752--223 in the light curve and spectral hardness.

\section{Observations and Results}

The MAXI carries two scientific instruments: the GSC 
 and the Solid State Camera (SSC). The GSC consists of twelve one-dimensional position-sensitive
 proportional counters operated in the 2--20 keV range, while 
  the SSC is composed of 32 X-ray CCD cameras with an energy range of 0.5--12 keV. 
 The GSC observes two different directions (horizontal and zenithal direction) with 
  an instantaneous field of view of 3$^{\circ}\times$160$^{\circ}$ each covered by six cameras.
  It covers 70 percent of the whole sky in every orbit 32 times per day at maximum. 
  Both instruments have been working properly in orbit, but four out of the twelve GSC cameras are turned off 
   due to discharges in the proportional counters.  See \citet{maxi} for more details of the MAXI.
In this paper, we use only the GSC data, which offer a larger sky coverage and effective area 
  than those of the SSC. 

\begin{figure*}[htbp]
  \begin{center}
    \FigureFile(90mm,90mm){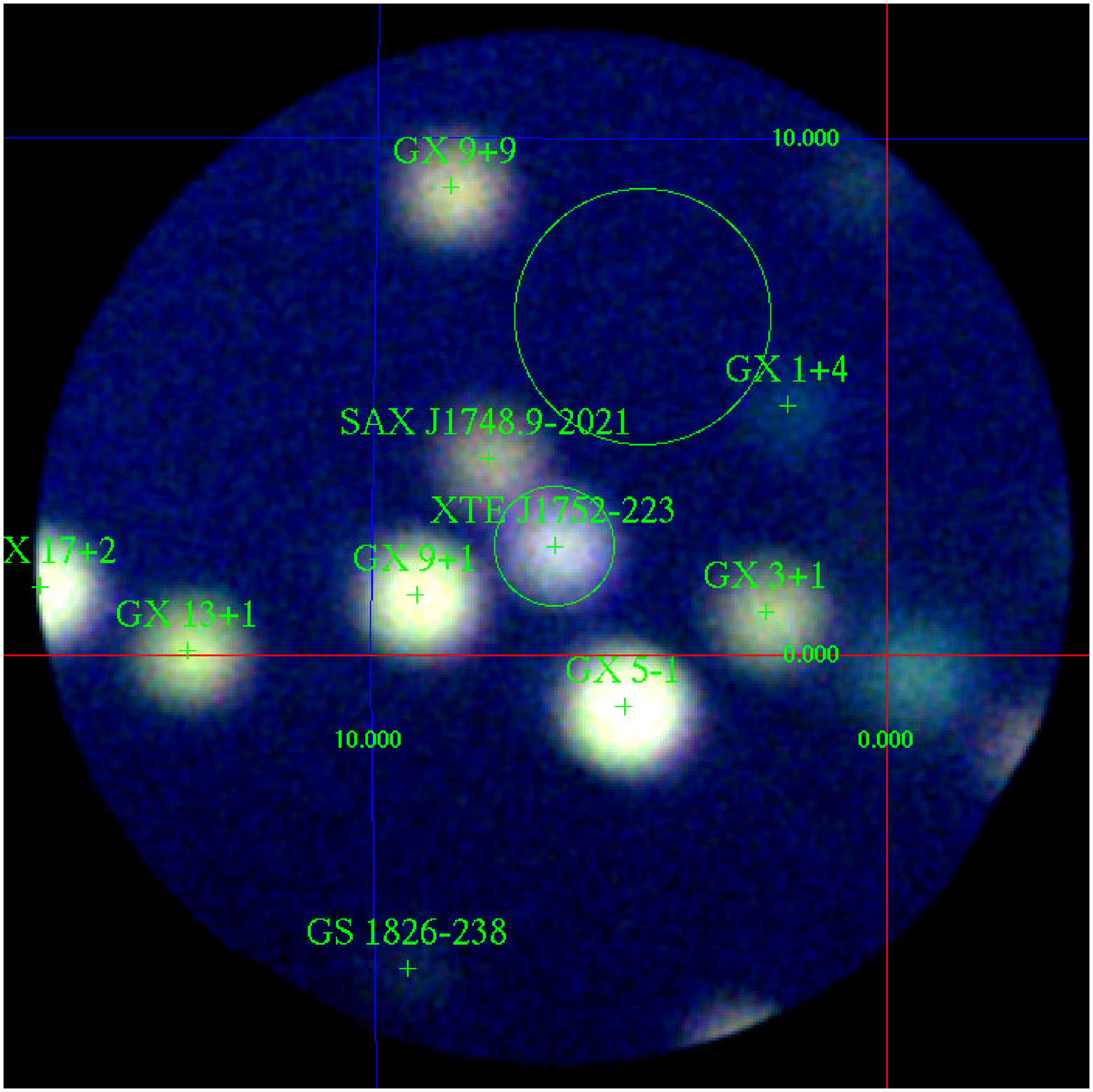}
    \FigureFile(90mm,90mm){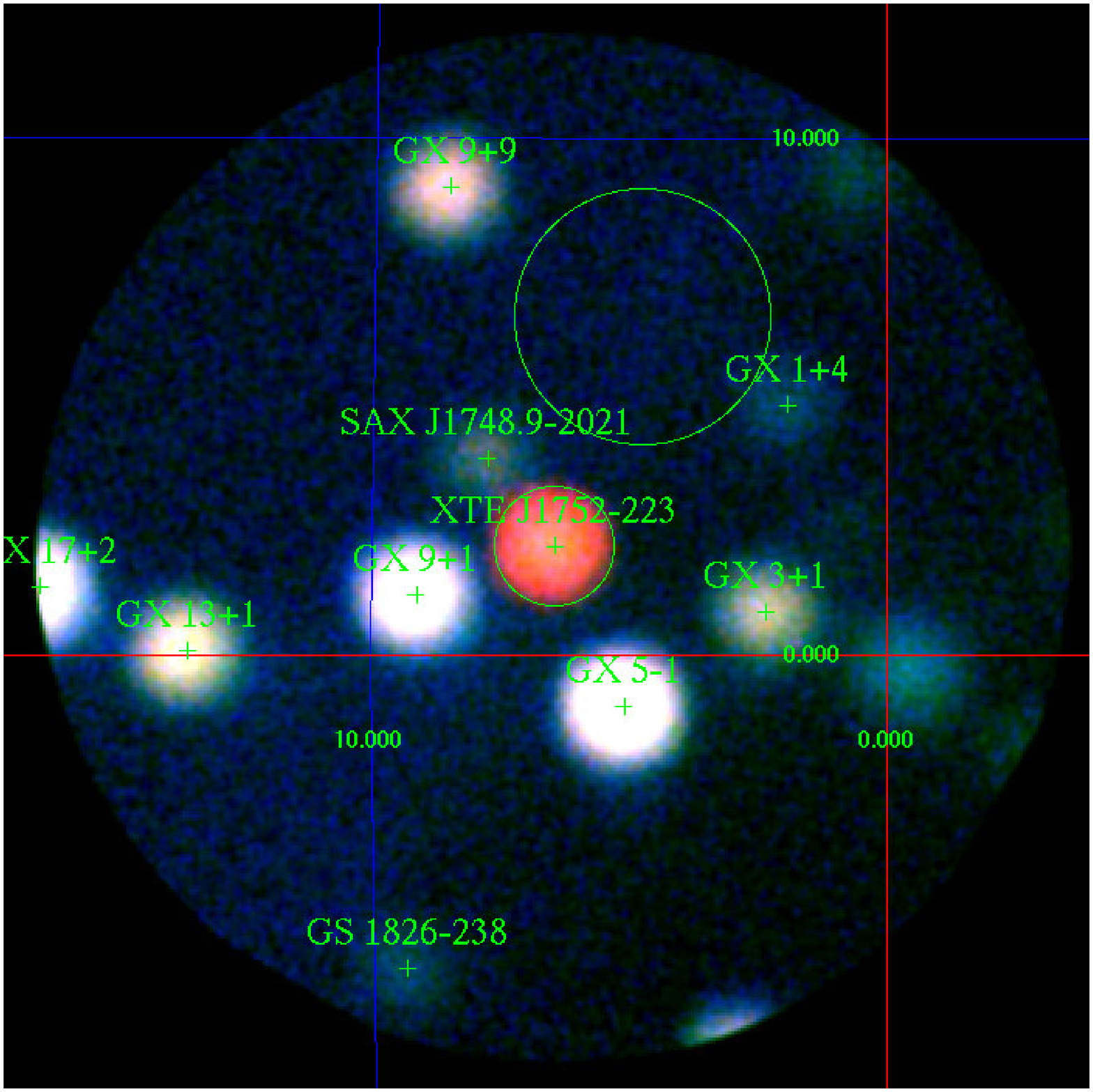}
  \end{center}
  \caption{MAXI/GSC false-color images during the period of December 20 to January 19 (upper panel) and January 20 to February 28 (lower panel). Red: 2--4 keV, 
Green: 4--10 keV, Blue: 10--20 keV. 10$^{\circ}$ radius area from XTE J1752--223 is extracted, and the source clearly reddens  because of the spectral transition. There are many bright X-ray sources near XTE J1752--223 with a few hundreds of mCrab such as GX 9+1 and GX 3+1. Source and background regions used for the light curve analysis
are shown by green circles on both images. }\label{fig1}
\end{figure*}

Figure \ref{fig1} shows two 1-month integrated X-ray sky images with a radius of 10 degrees centered
 on XTE J1752--223 in the 2--20 keV band. These false-color images are produced by 
superposing red (2--4 keV), green (4--10 keV), and blue (10--20 keV) images.  
 Data from all the counters were combined, and two images were made for 
   two separate periods (from December 20, 2009 to January 19, 2010, and 
January 20, 2010 to February 28, 2010). XTE J1752--223 is clearly detected in both GSC images. Because of its location
  near the Galactic Center (l, b)=(6.42, 2.11),  many bright X-ray sources 
 such as GX 9+1, GX 5--1, and GX 3+1 are visible in this field. 
The point spread function (PSF) of the MAXI/GSC is estimated at about 2.0 degrees (FWHM) in this energy range. 
The position of XTE J1752--223 determined by the GSC was consistent with the Swift/XRT localization.

\begin{figure}[htbp]
  \begin{center}
    \FigureFile(90mm,90mm){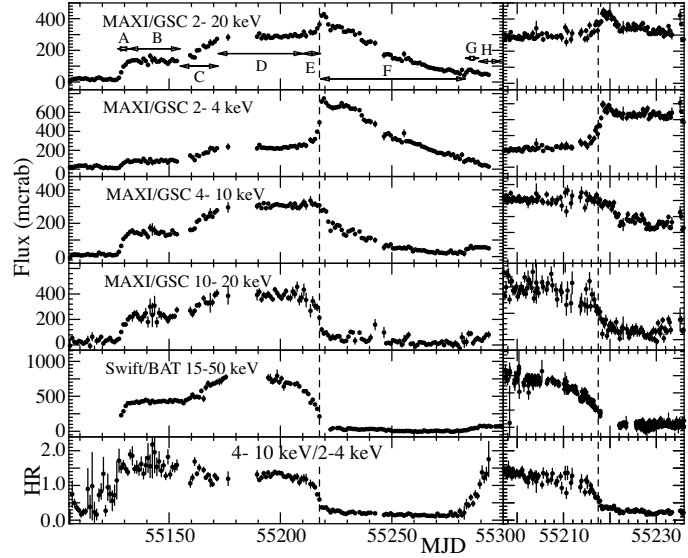}
  \end{center}
  \caption{The MAXI light curve of XTE J1752--223 in four energy ranges (2--20, 2--4, 4--10, and 10--20 keV from the top to the fourth panel)
 in comparison with Swift/BAT (15--50 keV: the fifth panel). The hardness ratio between the 2--4 keV and the 4--10 keV bands
is shown in the bottom panel. Right panel: The period in the left panel when the plasma ejection was detected (January 21 as shown by the dashed line) is expanded.}\label{fig2}
\end{figure}

 We calculated net source counts by subtracting the background from the on-source data.
 To avoid the source contamination from the nearby bright sources GX 9+1 (2.82$^{\circ}$ away) and SAX 
 J1748.9--2021 in the globular cluster NGC 6440 (\cite{ngc6440}; 2.13$^{\circ}$ away),  
 we carefully selected the source and background regions as circles of 1.17-degree and  2.48-degree radius, 
 respectively (see Figure \ref{fig1}). The extracted source counts were 
normalized by dividing by a total exposure 
 (in unit of cm$^2$s) obtained with a time integral of the collimator effective area.  
 The Crab Nebula  was analyzed in the same way, and the count rates of XTEJ 1752--223 were normallized by
 the Crab unit (1.30, 1.25, and 0.32 counts cm$^{-2}$ s$^{-1}$ in the 2--4, 4--10, and 10--20 keV range). 
 The 1-day averaged light curve of XTE J1752--223 thus obtained is shown in Figure \ref{fig2}. 
 The GSC scans were performed approximately  10 times per day.
 A data gap from December 7th to 23th was due to  solar avoidance limitation (the current solar  
    protection is set at above 4$^{\circ}$ from the Sun). 

 The X-ray light curve shows the following notable features. There are two initial plateau phases that 
 lasted for $\sim$25 days and $\sim$40 days, respectively, 
and the spectral hardness between them is slightly different: 
  the second phase (Phase D in Figure \ref{fig2}) has a softer spectrum than the first one 
  (Phase B). The source could not be observed during phase D by any other X-ray instruments except for RXTE/ASM and Swift/BAT
  due to the Sun constraints. The time scales of the rising phases A and C were about 5 days and 15 days, respectively.
   Another notable feature  is the anti-correlated behavior around the peak: 
  The 2--4 keV flux rapidly increased after January 20, 2010 (MJD: 55216) while 
   fluxes in the other two bands and the Swift/BAT 15--50 keV flux decreased (Phase E for $\sim$ 9 days). 
  The anti-correlation between the soft and hard bands indicates a spectral transition.
  This fact means that it took a very long time of $\sim$ 90 days for the source to complete the transition after 
  the onset of outburst. The 2--20 keV flux reached a peak on January 23, 2010 
  (MJD: 55219) at about the 430 mCrab level, then gradually decreased
  (Phase F). The emissions during Phase F are dominated by the soft bands ($<4$ keV) (see also the lower panel of 
  Figure \ref{fig1}). Around March 28 (MJD: 55283), hard X-rays and the hardness ratio again 
  increase (Phase G and H). The outburst continues at the time of writing this paper (the end of May 2010), and the 
  current duration for the outburst is about 210 days.

\begin{figure}[htbp]
  \begin{center}
   \rotatebox{90}{
    \FigureFile(90mm,90mm){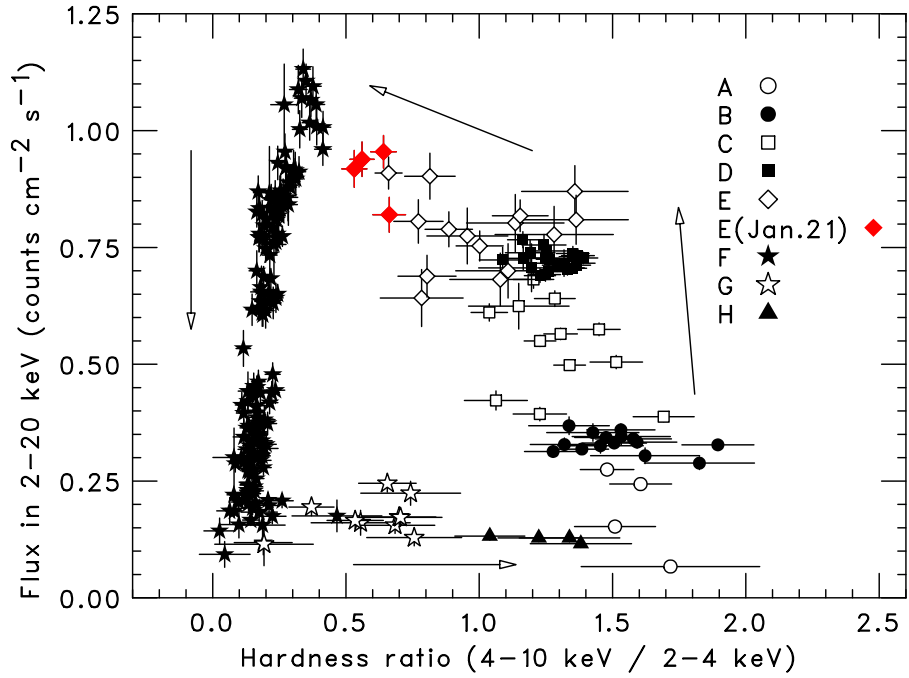}}
  \end{center}
  \caption{A  hardness--intensity diagram during the outburst of XTE J1752--223. Data points during the six phases (A to H) are shown by the different symbols. 6 hours integrations are used in Phase E, F and G, while one-day integrations are used for the other phases. The spectral evolution generally moves in the direction shown by arrows. Filled diamonds show the data when increased radio emissions, presumably jets, were detected on 2010 January 21. %The source is expected to go back to the low/hard state in the near future.
 }\label{fig3}
\end{figure}

 Figure \ref{fig3} shows a hardness-intensity diagram for the currently on-going outburst in XTE J1752--223.  
 Because of the anti-correlated behavior between the soft and hard X-rays during Phases E and G, 
  the data points track the Q-shaped curve, i.e. hysteresis behavior,  as seen in other BHCs (\cite{qcurve}). 
 The two plateau phases B and D correspond to the LHS, while the softer phase F corresponds with the HSS. 
 Thus, the state transition from the LHS to the HSS can be clearly identified with the MAXI/GSC data alone.   
 A color-color diagram is also shown in Figure \ref{fig4}. A line is plotted for a power-law spectral 
   model with a photon index ($\Gamma$) ranging from 1.0 to 5.0 modified by the absorption 
  (equivalent hydrogen column density N$_{\rm H}$=0.72$\times$10$^{22}$cm$^{-2}$; \cite{xte_result}).
  Note that the current GSC energy response gives systematic uncertainties of about 0.5$\times10^{22}$ cm$^{-2}$ and 0.1 
  in $N_{\rm H}$ and $\Gamma$, respectively. As the outburst proceeds, the $\Gamma$ continuously 
  varies from $\sim$1.5 to $\sim$5.0. The spectral shape during Phases B+D and F are 
  is approximated by a power-law with photon indices of 1.5--2.0 and above 3.5, respectively.
  These estimated photon indices are roughly consistent with typical values found from BHCs in the LHS (Phase B+D) and the HSS (Phase F).
  The power-law index in Phase E shows an intermediate value (2.0--3.0) between these seen in the LHS and the HSS.   
  An RXTE-pointed observation carried out on January 19, 2010 (MJD: 55215) during Phase E shows that 
   the source was at least in the IMS (\cite{xte_statetran}).
  Detailed spectral parameters and their evolution in this outburst will be reported in a forthcoming paper 
  (Yamaoka et al.\ 2010, in preparation). 
 
\section{Discussions}

  We have presented the first results on the black hole candidate XTE J1752--223 
  in the current outburst from the MAXI/GSC on the ISS. Judging from the light curve and spectral hardness, 
  we have found that the source was initially in the LHS
  with two long plateau phases (B and D). After the long stay ($\sim$90 days) in the LHS, 
  it exhibited a spectral transition via the IMS (Phase E) into the HSS around January 22, 2010 (Phase F).   
  The source came back to the LHS around April 3, 2010 (Phase H), and the source flux is gradually 
   declining toward the quiescent state.

\begin{figure}[htbp]
  \begin{center}
   \rotatebox{90}{
    \FigureFile(90mm,90mm){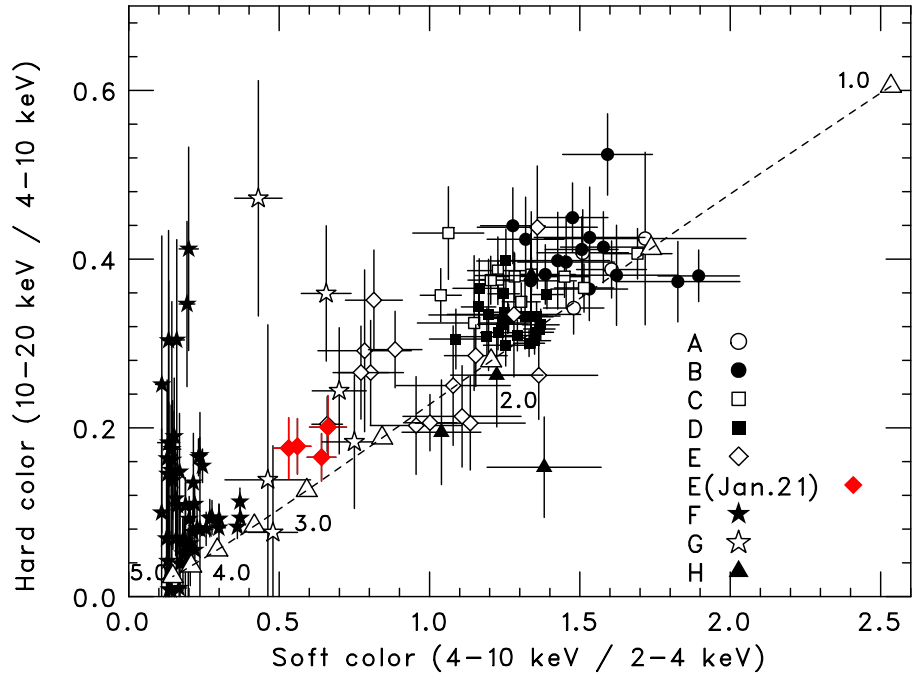}}
  \end{center}
  \caption{A color-color diagram during the  outburst of XTE J1752--223. The same symbols are used in Figure \ref{fig3}. The dashed line represents the spectral shapes of an absorbed 
  power-law with a photon index ranging from 1.0 to 5.0. The open triangles correspond to photon indices from  1.0 to  5.0 (0.5 step),  
from upper-right to lower-left.\label{fig4}}
\end{figure}

  It should be noted that a relatively high level of radio emission was detected by the Australia Telescope Compact Array 
  (ATCA) on January 21, 2010 (\cite{jet}), just corresponding to the beginning of the spectral transition from the LHS to 
  the HSS. The GSC spectrum on this date is roughly represented by a power-law with a 
   photon index of 2.5--3.0 (see Figure \ref{fig4}), suggesting that the source was 
   probably in the IMS.  Later, two separate radio sources were identified with high resolution 
  imaging by the European VLBI Network (e-EVN)  (\cite{jet2}). If we interpret it as plasma ejection, 
  it is consistent with the scenario that large-scale jets are produced in association with the 
  spectral transition from the LHS to the HSS. Another important result is that the luminosities during the transition from
  the HSS to the LHS, and from the LHS to the HSS were not the same: the hard-to-soft transition happened at a 2--20 keV flux higher by a 
   factor of about 7 than that at the soft-to-hard transition. Such hysteresis behavior was first noticed for 
   GS 1124--68 and GX 339--4 (\cite{hysteresis},) 
   and has been well established for many BHCs. It means that the spectral transition is not controlled       
   by the mass accretion rate alone, but depends on its history. 
  Most of the BHCs show a rapid flux rise within a few days that causes the near-instantaneous
   spectral transition (\cite{statetrans_class}), and then their fluxes decay with an e-folding time of several ten days into 
   quiescence (some BHCs show a secondary flare). It is known that the spectral state begins 
   with the initial LHS, then evolves through the VHS/IMS, HSS, and LHS during an outburst. However, about 8 BHCs 
   such as XTE J1118+480 and GRO J0422+32 remained in the LHS throughout their previous outbursts (\cite{lowhard_all}). 
   The very long duration of the initial LHS in XTE J1752--223 and 
   the two long plateau phases are rather uncommon for the BHC outbursts. 
  Meyer-Hofmeister (\cite{lowhard_orbit}) points out that systems with
  short orbital periods like XTE J1118+480 (4.1 hours) and GRO J0422+32
  (5.1 hours) tend to have low peak luminosities in their outbursts (and
  hence do not show state transition) because smaller accretion disks
  have less accumulated matter. The long LHS observed in XTE J1752--223
  might be related with a short orbital period. From our observations,
  however, XTE J1752-223 also resembles GX 339-4, which remained in the
  LHS for a long period before transition into the HSS (\cite{review}).
  GX 339-4 has a relatively long orbital period (1.7
  days) among BHCs categorized in low-mass X-ray binaries, indicating a
  larger accretion disk size. In this case, the unusually long duration
  of the LHS before the state transition would result from gradual
  change in the accretion rate. It is still unclear what causes the slow
  increase of the accretion rate in contrast with other X-ray
  novae. Therefore, the determination of orbital period of
  XTE J1752--223 should be important to clarify the nature of the long LHS
  period.
   
  Let us explain the overall picture of the light curve of the XTE J1752-223 outburst. 
  We assume the generally accepted disk truncation scenario in the LHS 
  never though some other scenarios have been proposed for the LHS, e.g. \cite{iscoLHS}, where the
  innermost radius of the disk ($R_{\rm in}$) is basically determined by the balance between the accretion rate and gas
  evaporation rate  (\cite{evap1}; \cite{evap2}; \cite{review2}).
  In this scenario, the LHS is the state where the accretion disk is truncated at a larger radius than
  the Innermost Stable Circular Orbit (ISCO =3 Schwarzschild radii for a non-rotating BH), 
  due to a low accretion rate. As the accretion rate increases ( thin begins in an outer part of the accretion disk), 
  $R_{\rm in}$ decreases. When the radius decreases, the corona around the disk
  cools down via efficient inverse Compton scattering of the soft disk photons and exhibits softer 
   X-ray spectra. Thus, the two phases in the LHS (B and D) can be interpreted with a slightly higher 
   accretion rate in Phase D than that in Phase B. Furthermore, when the accretion rate
 eventually exceeds a critical rate required for the state transition, the gas evaporation will not be 
  strong enough to truncate the disk. Thus, $R_{\rm in}$ reaches the ISCO, and the intense thermal 
  radiation from the accretion disk, i.e. ultra-soft component, appears in soft X-rays (Phase F in the HSS).
 The disk truncation scenario does not predict any time evolution of $R_{\rm in}$ during the HSS.  
 As seen in the light curve with a very smooth decay during the HSS, the local accretion rate smoothly 
   decreases with time. When it falls below the critical rate again, 
   the source comes back to the LHS (Phase H). 

  The significant increase mainly in the 2--4 keV band suggests that the disk innermost temperature ($kT_{\rm in}$) 
   is relatively low. Preliminary fits of the MAXI GSC data with the disk blackbody model (\cite{mcd}) during 
   Phase F revealed that $kT_{\rm in}$ was 0.4--0.7 keV including a systematic uncertainty 
   of 0.1 keV. The Swift XRT observation taken on February 4, 2010 also showed that the blackbody 
   temperature is about 0.5 keV (\cite{lowtin}). This temperature is lower than those of 
    other BHCs such as GRO J1655--40 and XTE J1550--564 whose $kT_{\rm in}$ in the HSS exceeds 1 
   keV (\cite{disk}). The vary of $kT_{\rm in}$ is determined by how much the accretion rate exceeds  
   the critical rate, hence a low $kT_{\rm in}$ indicates that the accretion rate in the HSS is above the critical rate. 
  It is known that the spectral transition from the LHS to the HSS will occur at
     around 10--20\% Eddington luminosity, but with large uncertainties (\citet{translum}). 
  The k$T_{\rm in}$ for the Schwarzschild BH can be estimated to be  
  $\approx 1.2 {\rm keV} (M/10M_\odot)^{-1/4} (\eta/1.0)^{1/4} (h/1.7)$ (e.g., \citet{ulx}),
  where $h$ is the spectral hardening factor (\cite{hardening}), $\eta$ is the Eddington ratio ($L/L_{\rm E}$), 
  and $M$ is the mass of the central object. Assuming that the maximum $kT_{\rm in}$ is 0.6$\sim$0.7 keV and $\eta$ is 0.1$\sim$0.2, 
    the black hole mass can be roughly estimated at 8.6--32 M$_{\odot}$.  
  The GSC flux in the 2--10 keV range at the outburst peak (2010 January 23) was
   9.7$\times$10$^{-9}$ erg cm$^{-2}$ s$^{-1}$ with a $\sim$20\% absolute flux uncertainty. 
  Assuming that 1) the 2--10 keV flux is 21--28 \% of the bolometric flux based on the disk blackbody model, 
  and that 2) the Eddington luminosity is 1.5$\times$10$^{38} (M/M_\odot)$  erg s$^{-1}$, the distance to the source can be estimated as 5--10 kpc.
%   These parameters will be confirmed by future observations of this system.  

 The authors are indebted to many other members of the MAXI team and the ISS/MAXI operation team for their efforts during
the data acquisition and  the daily MAXI operation. 
 This research was partially supported by the Ministry of Education, Culture, Sports, Science and Technology (MEXT), Grant-in-Aid No.19047001, 20041008, 20540230, 
20244015 , 21340043, 21740140, 22740120, and Global-COE from MEXT ``The Next Generation of Physics, Spun from Universality and Emergence'' and 
``Nanoscience and Quantum Physics''. 
One of the authors (K. Y.) is grateful to Dr. T. Kawaguchi and Dr. P. Gandhi for valuable discussion. 
The authors also acknowledge the Swift/BAT team for making the hard X-ray transient monitor 
 results publicly available\footnote {http://heasarc.gsfc.nasa.gov/docs/swift/results/transients/}.

%\bigskip

%Acknowledgement should be placed at end of main text.
%(NOT after the Appendix.)

%%%
% See the manual for the detail.
%%%

\clearpage

\begin{thebibliography}{}
% Journals(e.g. A\&A,ApJ,AJ,NMRAS,PASP ...)
% Authors, Year, Journal, Vol#, Page#
% Journal Title Abbreviation >> http://www.asj.or.jp/pasj/Jabb.html
\bibitem[Brocksopp et al.(2001)]{lowhard_all} Brocksopp, C., Jonker, P. G., Fender, R. P., Groot, P. J., van der Klis, M., Tingay, S. J., 2001, MNRAS, 323, 517
\bibitem[Brocksopp et al.(2009)]{radio} Brocksopp, C., Corbel, S., Zunis, T., Fender, R.,  2009, ATel, 2278
\bibitem[Brocksopp et al.(2010a)]{jet} Brocksopp, C., Corbel, Tzioumis, T., Fender, R., Coriat, M.,  2010, ATel, 2400
\bibitem[Brocksopp et al.(2010b)]{jet2} Brocksopp, C., Yang, J., Corbel, S., Zunis, T., Fender, R.,  2010, ATel, 2438
\bibitem[Curran et al.(2010)]{lowtin} Curran, P. A., Evans, P. A., Still, M., Brocksopp, C., Done, C., 2010, ATel, 2424
\bibitem[Done, Gierli\'nski, \& Kubota (2007)]{review2} Done, C., Gierli\'Nani, M., Kubota, A., 2007, A\&AR, 15, 1 
\bibitem[Esin, McClintock, \& Narayan (1997)]{flow1} Esin, A. A., McClintock, J. E., \& Narayan, R., 1997, \apj, 489, 865
\bibitem[Fender(2006)]{review_jet} Fender, R., 2006, in Compact Stellar X-ray Sources, ed. W. H. G. Merwin \&M. van der KS (Cambridge: Cambridge Univ. Press), 381
\bibitem[Gierli\'nski \& Newton(2006)]{statetrans_class} Gierli\'nski, M., \& Newton, J., 2006, \mnras, 370, 837
\bibitem[Gierli\'nski \& Done(2004)]{disk} Gierli\'nski, M., \& Done, C., 2004, \mnras, 347, 885
\bibitem[Homan \& Belloni(2005)]{qcurve} Homan, J., \& Belloni, T., 2005, Sap\&SS, 300. 107
\bibitem[Liu et al.(2002)]{evap2} Liu, B. F., Mineshige, S., Meyer, F., Meyer-Hofmeister, E., Wac, T., 2002, \apj, 575, 117
\bibitem[Liu et al.(2007)]{iscoLHS} Liu, B. F., Taam, R. E., Meyer-Hofmeister, E., Meyer, F., 2007, \apj, 671, 695
\bibitem[Makishima et al.(2000)]{ulx} Makishima, K., et al., 2000, \apj, 535, 632 
\bibitem[Markoff et al.(2001)]{flow2} Markoff, S., Falcke, H., \& Fender, R., 2001, A\&A, 372, L25
\bibitem[Markwardt et al.(2009a)]{discovery} Markwardt, C. B., et al., 2009, ATel, 2258
\bibitem[Markwardt et al.(2009b)]{swiftxrt} Markwardt, C. B., Barthelmy, S. D., Evans, P. A., Swank, J. H., 2009, ATel, 2261
\bibitem[Matsuoka et al.(2009)]{maxi} Matsuoka, M., et al., 2009, \pasj, 61, 999
\bibitem[Matsuoka et al.(2010)]{maxi_first} Matsuoka M., et al. 2010, Proc. ``X-ray Astronomy 2009'', Bologna, Italy, Sep. 7-11, 2009, UP, eds. A. Comastri, M. Cappi, and L. Angelini in press.
\bibitem[McClintock \& Remillard(2006)]{review} McClintock, J. E., \& Remillard, R.A., 2006, in Compact Stellar X-ray Sources, ed. W.H.G. Merwin \&M. van der Klis (Cambridge: Cambridge Univ. Press), 157
\bibitem[Meyer, Liu, \& Meyer-Hofmeister(2000)]{evap1} Meyer, F., Liu, B.F., \& Meyer-Hofmeister, E., 2000, A\&A, 354, L67
\bibitem[Meyer-Hofmeister(2004)]{lowhard_orbit} Meyer-Hofmeister, E., 2004, A\&A, 423, 321
\bibitem[Mitsuda et al.(1984)]{mcd} Mistier, K., et al., 1984, \pasj, 36, 741
\bibitem[Miyamoto et al.(1995)]{hysteresis} Miyamoto et al., 1995, \apj, 442, L13
\bibitem[Munoz-Darias et al.(2010)]{xte_result} Munoz-Darias, T., Motta, S., Parer, D., Belloni, T. M., Campana, S., Bhattacgarta, D., 2010, arXiv:1003.0477
\bibitem[Nakahira et al.(2009)]{maxiatel1} Nakahira, S. et al., 2009, ATel, 2259
\bibitem[Negoro et al.(2010)]{maxiatel2} Negoro, H. et al., 2010, ATel, 2396
\bibitem[Patruno et al.(2010)]{ngc6440} Patna, A. et al., 2010, ATel, 2407
\bibitem[Shaposhnikov et al.(2009)]{xte_spec} Shaposhnikov, N.,  Markwardt, C. B., \& Swank, J. H., 2009, ATel, 2269
\bibitem[Shaposhnikov(2010)]{xte_statetran} Shaposhnikov, N., 2010, ATel, 2391
\bibitem[Shimura \& Takahara (1995)]{hardening} Shimura, T., \& Takahara, F., 1995, \apj, 445, 780
\bibitem[Torres et al.(2009a)]{opt} Torres, M. A. P., Jonker, P. G., Steeghs, D., Yan, H., Huang, J., Soderberg, A. M.,  2009, ATel, 2263
\bibitem[Torres et al.(2009b)]{opt+nir} Torres, M. A. P., Steeghs, D., Jonker, P. G., Thompson, I., Soderberg, A. M.,  2009, ATel, 2268
\bibitem[Ueno et al.(2010)]{maxi_first2} Ueno, S., et al., 2010, in proceedings of ``The Extreme sky: Sampling the Universe above 10 keV'' PoS(extremesky2009)011
\bibitem[Wu et al.(2010)]{translum} Wu, X. Y., Yu, W., Sun, L., Li, T. P., 2010, A\&A, 512, 32
\bibitem[Yuan, Cui, \& Narayan(2005)]{flow3} Yuan, F., Cui, W., \& Narayan, R., 2005, \apj, 620, 905
\end{thebibliography}
\end{document}